\documentclass[hyper]{JHEP} 
\usepackage{epsfig}
\newcommand\fverb{\setbox\pippobox=\hbox\bgroup\verb}
\newcommand\fverbdo{\egroup\medskip\noindent%
            \fbox{\unhbox\pippobox}\ }

\newcommand\fverbit{\egroup\item[\fbox{\unhbox\pippobox}]}
\newbox\pippobox
\title{Rotating Strings in AdS$_4\times$ CP$^3$ with B$_{\rm NS}$ holonomy}
\author{Kamal L. Panigrahi, Pabitra M. Pradhan and Pratap K. Swain \\
Department of Phyisics and Meteorology, \\Indian Institute of
technology Kharagpur,
Kharagpur-721 302, INDIA \\
E-mail: \email{panigrahi,ppabitra,pratap@phy.iitkgp.ernet.in}}
\abstract{We study solutions for rigidly rotating strings on
AdS$_4 \times$ CP$^3$ in the presence of $B_{\rm NS}$ holonomy
turned on over $CP^1 \subset CP^3$. We construct general solutions
for rotating strings with two and three angular momenta in $CP^3$
and discuss various limits corresponding to giant magnon and spike
like solutions.} \keywords{AdS-CFT correspondence, Bosonic
Strings}

\begin{document}

\section{Introduction}

The proposed ABJM theory \cite{Aharony:2008ug} has been
conjectured to be dual to M-theory on $AdS_4 \times  S^7/Z_k$ with
$N$ units of four-form flux, which for $k << N << k^5$ can be
compactified to type IIA theory on $AdS_4 \times CP^3$, where $k$
is the level of Chern-Simon (CS) theory with gauge group $U(N)
\times U(N)$. In continuation with this proposal, Aharony,
Bergman, and Jafferis (ABJ) \cite{Aharony:2008gk} identified a
further class of gauge-gravity duality with extended supersymmetry
namely, a three dimensional $N = 6$ superconformal CS theory with
a gauge group $U(M)_k \times {\overline{U(N)}}_{-k}$, with $k$
being the level of the CS theory, is dual to type IIA string
theory on $AdS_4\times CP^3$ with a two form B$_{NS}$ holonomy
turned on over $CP^1 \subset CP^3$. In the understanding of the
$AdS_4/CFT_3$ duality better \cite{Minahan:2008hf,
Nishioka:2008gz, Gaiotto:2008cg, Grignani:2008is, Gromov:2008qe,
Krishnan:2008zs, Berenstein:2008dc, McLoughlin:2008he}, the
semiclassical string states in the gravity side have been used to
look for suitable gauge theory operators on the boundary. In this
connection, the rigidly rotating strings with large angular
momentum and energy have been considered and in special limits
they correspond to the so called giant magnon \cite{Hofman:2006xt}
and single spike \cite{Kruczenski:2004wg} solutions. The
corresponding dual field theory with long trace operators with
large angular momentum and energy have also been studied in great
detail. The generalization of the so called magnon dispersion
relation in the presence of two and three angular momenta has also
been considered in both asymptotically AdS and non-AdS backgrounds
\footnote{see for example \cite{Bobev:2005cz, Ryang:2005pg,
Chu:2006ae, Kruczenski:2006pk, Kluson:2007qu, Ishizeki:2007we,
Bobev:2007bm, Dimov:2007ey, Lee:2008sk, David:2008yk, Lee:2008ui,
Ryang:2008rc, Abbott:2008qd, Suzuki:2009sc, Abbott:2009um}.}. More
recently a class of three spin spiky strings in ABJM model has
been studied and various interesting solutions has been obtained
in \cite{Giardino:2011dz}.

In this paper, we study the semi-classical string solutions in ABJ
model. In \cite{Bak:2008vd}, it has been shown that ABJ theory
also has an integrable structure in planner limit. The rigidly
rotating strings corresponding to giant magnon and single spike
solution along with the finite size correction has been studied
earlier. However the subspace that was considered in
\cite{Jain:2008mt} was simply $R_t \times S^2$ and hence the
solutions correspond to one angular momentum. We wish to study the
two spin and three spin solutions in the background of AdS$_4
\times$ CP$^3$ in the presence of NS-NS B-field. We have shown
that the corresponding solutions obtained in \cite{Ryang:2008rc,
Giardino:2011dz} will get contributions due to the presence of the
B-field in a very natural way. Further we construct a class of
folded string solution in ABJ model and obtain a relation among
various conserved charges. The rest of this paper is organized as
follows. In section-2, we construct rigidly rotating strings in
the background of $AdS_4 \times CP^3$ with NS-NS B-field with two
angular momenta on CP$^3$. We found two classes of solutions
corresponding to giant magnon and spiky strings. We compute all
the conserved charges and found out the dispersion relation among
them. In section-3, we generalize this solution to three angular
momenta on CP$^3$. Section-4 is devoted to the study of a class of
folded string solutions in ABJ model. Finally in section-5 we
conclude with some remarks.

\section{Spiky strings with two angular momenta}
In this section we study two spin rigidly rotating strings in ABJ
model. The metric and NS-NS fields ($B_{\mu\nu}$) for the ABJ
model, that is the background of $AdS_4 \times CP^3$ with $B_{\rm{
NS}}$ holonomy is given by
\begin{eqnarray}
ds^2 &=& \frac{R^2}{4}\Big(-\cosh^2\rho dt^2 + d\rho^2 +
\sinh^2\rho d\Omega_2^2\Big) + R^2\Big[d\xi^2 + \cos^2\xi
\sin^2\xi\Big(d\psi + \frac{1}{2}\cos\theta_1d\phi_1 \cr & \cr
&&-\frac{1}{2}\cos\theta_2d\phi_2\Big)^2 +
\frac{1}{4}\cos^2\xi(d\theta_1^2 + \sin^2\theta_1d\phi_1^2) +
\frac{1}{4}\sin^2\xi(d\theta_2^2 + \sin^2\theta_2d\phi_2^2) \Big],
\cr &\cr B_{NS} &=& -B\sin2\xi ~ d\xi \wedge d\psi -\frac{B}{2}
\sin2\xi\cos\theta_1 ~ d\xi \wedge d\phi_1 +
\frac{B}{2}\sin2\xi\cos\theta_2d\xi \wedge d\phi_2
 \cr & \cr &&- \frac{B}{2}\cos^2\xi\sin\theta_1 d\theta_1 \wedge d\phi_1
 -\frac{B}{2}\sin^2\xi\sin\theta_2 d\theta_2 \wedge d\phi_2,
\nonumber \\ \label{1}
\end{eqnarray}
where $0 \leq \xi \leq \frac{\pi}{2}$, $-2\pi \leq \psi \leq 2\pi$
and ($\theta_i, \phi_i$) are coordinates on two sphere. The radius
$R$ is related to 't Hooft coupling constant $\lambda$ as $R^2 =
2^{\frac{5}{2}}\pi\sqrt{\lambda}$. We are interested in studying
the string solutions in a complimentary subspace of $AdS_4 \times
CP^3$ for which we take the choice of coordinate in (\ref{1}) as
$\rho = 0$ and $\theta_{1,2} = \frac{\pi}{2}$   and $\phi_1 =
\phi_2 =\phi$. With the above choice the metric and B field
reduces to
\begin{eqnarray}
ds^2 &=& - \frac{1}{4}R^2 dt^2 + R^2[ d\xi^2 + \cos^2\xi \sin^2\xi
d\psi^2 + \frac{1}{4} d\phi^2], \cr & \cr B_{NS} &=& -B \sin 2\xi
~ d\xi \wedge d\psi. \nonumber \\ \label{2}
\end{eqnarray}
We are interested in studying the solutions of rigidly rotating
strings in the above background. We start by writing down the
Polyakov action of the fundamental string in the background of
(\ref{2}), i.e.
\begin{eqnarray}
S &=& T \int ~ d\sigma d\tau \Big[ - \frac{1}{4}({\dot{t}}^2 -
{t^{\prime}}^2) + \dot{\xi}^2 - {\xi^{\prime}}^2 + \cos^2\xi
\sin^2\xi(\dot{\psi}^2 - {\psi^{\prime}}^2) +
\frac{1}{4}(\dot{\phi}^2 - {\phi^{\prime}}^2) \Big] \cr & \cr &&+
\frac{B}{4\pi} \int ~ d\sigma d\tau \Big[ 2\sin 2\xi
(\dot{\psi}\xi^{\prime} - \dot{\xi} \psi^{\prime})\Big], \nonumber
\\ \label{3}
\end{eqnarray}
where $ T = \sqrt{2\lambda}$ . The equations of motion derived
from the above action (\ref{3}) are
\begin{eqnarray}
\sin^2 2\xi({\psi}^{\prime \prime} - \ddot{\psi}) +
\partial_{\xi}(\sin^2 2\xi)
(\xi^{\prime}\psi^{\prime} - \dot{\xi}\dot{\psi}) &=& 0,
 \cr & \cr
{\xi}^{\prime \prime} - \ddot{\xi} -\frac{1}{8}
\partial_{\xi}(\sin^2 2\xi)
({\psi^{\prime}}^2 - {\dot{\psi}}^2) &=& 0,
 \cr & \cr
{\phi}^{\prime \prime} - \ddot{\phi} &=& 0,
 \cr & \cr {t}^{\prime
\prime} - \ddot{t} &=& 0.
\nonumber \\ \label{4}
\end{eqnarray}

To study the rotating string with two angular momenta we take the
following ansatz:
\begin{eqnarray} t = \kappa \tau,\> \psi = \omega \tau + f(y), \>
\xi = \xi(y), \> \phi = \nu \tau \ ,
\end{eqnarray}
where we have defined $y = a\sigma + b\tau$. On the other hand,
the two Virasoro constraints derived from the action (\ref{3}) are
given by
\begin{eqnarray}
T_{\tau \tau} + T_{\sigma\sigma} = {\xi_y}^2 - \frac{1}{4}
\frac{{\kappa}^2 - \nu^2} {a^2 + b^2} + \frac{1}{4}\sin^2 2\xi
\big({f_y}^2 + \frac{\omega^2 + 2b\omega f_y}{a^2 + b^2}\big) &=&
0 , \cr & \cr T_{\tau \sigma} = {\xi_y}^2 + \frac{1}{4}\sin^2
2\xi \big({f_y}^2 + \frac{\omega f_y}{b}) &=& 0. \nonumber \\
\label{6}
\end{eqnarray}
Where $\xi_y$ and $f_y$ denote the partial derivatives of $\xi$
and $f$ with respect to $y$. The difference of these two Virasoro
constraints gives the following relations among various constants,
\begin{equation}
C = - \frac{b}{\omega}(\kappa^2 - \nu^2) \ , \label{7}
\end{equation}
where $C$ is the integration constant coming from solving the
equation of motion for $\psi$ in (\ref{4}) and the addition gives
\begin{equation}
\xi_y = \pm \frac{a \omega}{2(a^2 - b^2)\sin 2\xi} \sqrt{(\cos^2
2\xi_+ - \cos^2 2\xi) (\cos^2 2\xi - \cos^2 2\xi_-)} , \nonumber
\\ \label{9}
\end{equation}
where $\cos^2 2\xi_{\pm} =$
\begin{eqnarray}
\frac{2 a^2 \omega^2 -(\kappa^2 - \nu^2)(a^2 + b^2)}{2 a^2
\omega^2} \Big[ 1\pm \sqrt{1 - \frac{(\kappa^2 - \nu^2 -
\omega^2)((\kappa^2 - \nu^2)a^2 b^2 -a^4 \omega^2)}{(2a^2 \omega^2
- (\kappa^2 - \nu^2)(a^2 + b^2))^2}}\Big].
\nonumber \\
 \label{10}
\end{eqnarray}
The conserved quantities associated with symmetry of the action
(\ref{3}) are
\begin{eqnarray}
E &=& \pm 2T \frac{\kappa( a^2 - b^2)}{a^2 \omega}
\int_{\xi_+}^{\xi_-} ~d\xi \frac{\sin 2\xi}{\sqrt{(\cos^2 2\xi_{+}
- \cos^2 2\xi) ( \cos^22\xi - \cos^2 2\xi_{-})}},
 \cr & \cr
J_{\phi} &=& \pm 2T \frac{\nu( a^2 - b^2)}{a^2 \omega}
\int_{\xi_+}^{\xi_-}
 ~d\xi \frac{\sin 2\xi}
{\sqrt{(\cos^22\xi_{+} - \cos^2 2\xi) ( \cos^2 2\xi - \cos^2
2\xi_{-})}},
 \cr & \cr J_{\psi} &=& \pm 2T \frac{a^2 - b^2}{a^2
\omega}
 \int_{\xi_+}^{\xi_-} ~d\xi \frac{\sin^3 2\xi}
{\sqrt{(\cos^2 2\xi_{+} - \cos^2 2\xi)( \cos^2 2\xi - \cos^2
2\xi_{-})}} \cr & \cr && \left[ \omega + \frac{b^2}{\omega(a^2 -
b^2)}\left(\omega^2 - \frac{\kappa^2 - \nu^2}{\sin^2
2\xi}\right)\right]
 - \frac{B}{2\pi} \int_{\xi_+}^{\xi_-} ~ d\xi ~ 2 \sin 2\xi.
\nonumber \\ \label{11}
\end{eqnarray}
Taking an infinite volume limit as $\xi_{-} = \frac{\pi}{4}$, then
from (\ref{10}), we have two conditions as $(i)~ \kappa^2 - \nu^2
= \omega^2 $ and $(ii)~ \kappa^2 - \nu^2 =
\frac{a^2}{b^2}\omega^2$. Using the condition (i) we get
\begin{equation}
\kappa E - \nu J_{\phi} - \omega J_{\psi} = \left( T +
\frac{B}{2\pi}\right) \omega \cos 2\xi_{+}, \nonumber \\
\label{14}
\end{equation}
and the dispersion relation as
\begin{equation}
\sqrt{ E^2 - {J_{\phi}}^2} - J_{\psi} = \left( T +
\frac{B}{2\pi}\right) \cos 2\xi_{+}, \nonumber \\ \label{15}
\end{equation}
which is a giant magnon like dispersion relation. Using the
condition (ii), we get a spike like dispersion relation same as
\cite{Ryang:2008rc}, which is
\begin{equation}
\sqrt{E^2 - {J_{\phi}}^2} - \frac{1}{2} T \delta\psi = T
\left(\frac{\pi}{2} - 2 \xi_{+}\right), \nonumber \\ \label{16}
\end{equation}
with $J_{\psi} = -\left(T + \frac{B}{2 \pi}\right) \cos 2\xi_{+}$.

\section{Rotating strings with three angular momenta}
In this section, we generalize the above two spin string solution
to three spin string solution in another class of subspace of
$CP^3$. Here we take the choice of coordinates in (\ref{1}) as
$\rho = 0$ and $\theta_{1,2} = \frac{\pi}{2}$. Thus the new metric
and B-field look like
\begin{eqnarray}
ds^2 &=& -\frac{R^2}{4} dt^2 + R^2 \Big[d\xi^2 +
\cos^2\xi\sin^2\xi d\psi^2 + \frac{1}{4}\cos^2\xi d\phi_1^2 +
\frac{1}{4}\sin^2\xi d\phi_2^2\Big], \cr & \cr B_{NS} &=&
-B\sin2\xi ~ d\xi \wedge d\psi. \nonumber \\ \label{17}
\end{eqnarray}
We start by writing down the Polyakov action of the fundamental
string in the background (\ref{17}) as
\begin{eqnarray}
S &=& -T\int d\sigma d\tau \Big[ -\frac{1}{4} ({t^{\prime}}^2 -
\dot{t}^2) + ({\xi^{\prime}}^2 - \dot{\xi}^2) +
\cos^2\xi\sin^2\xi({\psi^{\prime}}^2 - \dot{\psi}^2) + \cr & \cr
&&+ \frac{1}{4}\cos^2\xi({\phi_1^{\prime}}^2 - \dot{\phi_1}^2) +
\frac{1}{4}\sin^2\xi({\phi_2^{\prime}}^2 - \dot{\phi_2}^2)\Big]
\cr & \cr &&- \frac{1}{2\pi}\int d\sigma d\tau ~ B\sin 2\xi ~
(\dot{\xi}\psi^{\prime} - \xi^{\prime}\dot{\psi}). \nonumber \\
\label{18}
\end{eqnarray}

We choose the following ansatz
\begin{eqnarray}
t=\kappa\tau, ~~~ \xi = \xi(y), ~~~ \psi = \omega\tau + f(y), ~~~
\phi_{i=1,2} = \nu_i\tau + g_i(y) , \end{eqnarray} where $y =
a\sigma + b\tau$ and $a, b, \kappa, \omega$ ~ and $\nu_i$ are
constants. Then the equations of motion derived from the above
action (\ref{18}) is same as \cite{Giardino:2011dz} as the
B-fields are appearing in such a way that their contributions to
equations of motion get canceled. The first Virasoro constraint is
\begin{eqnarray}
T_{\tau\sigma} &=&  a  b {\xi_y}^2 + \cos^2\xi\sin^2\xi(\omega a
f_y +
 a  b {f_y}^2) + \frac{1}{4}\cos^2\xi
\Big(\nu_1 a g_{1y} +
 a  b {g_{1y}}^2\Big) \cr & \cr &&+
\frac{1}{4}\sin^2\xi\Big(\nu_2 a  g_{2y} +
 a  b {g_{2y}}^2\Big) = 0,
\nonumber \\ \label{20}
\end{eqnarray}
and the second virasoro constraint is
\begin{eqnarray}
T_{\tau\tau} + T_{\sigma\sigma} &=& - \frac{1}{4}{\kappa}^2 + ({ a
}^2 + { b }^2){\xi_y}^2 + \cos^2\xi\sin^2\xi\Big[{\omega}^2 + ({ a
}^2 + { b }^2){f_y}^2 + 2\omega  b f_y\Big] \cr & \cr &&+
\frac{1}{4}\cos^2\xi\Big[{\nu_1}^2
 + (a^2 + b^2){g_{1y}}^2
 + 2\nu_1  b g_{1y}\Big]
\cr & \cr &&+ \frac{1}{4}\sin^2\xi\Big[{\nu_2}^2 + ({ a }^2 + {
b}^2){g_{2y}}^2 +
 2\nu_2  b g_{2y}\Big] = 0.
\nonumber \\ \label{21}
\end{eqnarray}
These Virasoro constraints are same as that given in
\cite{Giardino:2011dz}. Eliminating ${\xi_y}^2$ from (\ref{20})
and (\ref{21}) and using results of equation of motion we get the
following relation on virasoro constraints
\begin{equation}
\frac{{\kappa}^2 b }{{ a }^2 - { b }^2} + 4\omega A_{\psi} + \nu_1
A_1 + \nu_2 A_2 = 0, \nonumber \\ \label{22}
\end{equation}
where $A_{\psi}$, $A_1$ and $A_2$ are integration constants coming
from solutions of equation of motions of $\psi$, $\phi_1$ and
$\phi_2$ respectively. The conserved conjugate momenta that are
associated with the coordinates $t, \psi, \phi_1$ and $\phi_2$
remains same as in case of ABJM model given in
\cite{Giardino:2011dz} except the conserved quantity that
associated with coordinate $\psi$ which gets extra contribution
due to the presence of B-field. The conserved momenta are
\begin{eqnarray}
E &=& 2T \frac{\kappa}{\omega}\frac{{ a }^2 - { b }^2} { a }
\int_{\xi_+}^{\xi_-} ~d\xi \frac{\sin 2\xi}{\sqrt{(\cos^2 2\xi_{+}
- \cos^2 2\xi) ( \cos^22\xi - \cos^2 2\xi_{-})}} ,\cr & \cr
J_{\phi_1} &=& \Big( b  A_1 + \frac{1}{2}\frac{\omega_1 { a }^2}{{
a }^2 - { b }^2}\Big)\frac{E}{\kappa} \mp
\frac{T}{2}\frac{\omega_1}{\omega} \int_{\xi_+}^{\xi_-} ~d\xi
\frac{\cos 2\xi \sin 2\xi}{\sqrt{(\cos^2 2\xi_{+} - \cos^2 2\xi) (
\cos^22\xi - \cos^2 2\xi_{-})}} ,\cr & \cr J_{\phi_2} &=& \Big( b
A_2 + \frac{1}{2}\frac{\omega_2 { a }^2}{{ a }^2 - { b
}^2}\Big)\frac{E}{\kappa} \mp \frac{T}{2}\frac{\omega_2}{\omega}
\int_{\xi_+}^{\xi_-} ~d\xi \frac{\cos 2\xi \sin
2\xi}{\sqrt{(\cos^2 2\xi_{+} - \cos^2 2\xi) ( \cos^22\xi - \cos^2
2\xi_{-})}} ,\cr & \cr J_{\psi} &=& \Big(4 b A_{\psi} +
\frac{\omega { a }^2}{{ a }^2 - { b }^2}\Big)\frac{E}{\kappa} - T
\int_{\xi_+}^{\xi_-} ~d\xi \frac{\cos^2 2\xi \sin
2\xi}{\sqrt{(\cos^2 2\xi_{+} - \cos^2 2\xi) ( \cos^22\xi - \cos^2
2\xi_{-})}} \cr & \cr &&- \frac{B}{2\pi}~\int_{\xi_+}^{\xi_-} d\xi
~ 2 \sin 2\xi,
\end{eqnarray}
where
\begin{eqnarray}
\cos^2 2\xi_{\pm} &=& 1 + \frac{1}{2}\Big( \frac{\nu^2 -
\kappa^2}{\omega^2} - \frac{b^2 \kappa^2}{a^2 \omega^2}\Big) \pm
\frac{1}{2 \omega^2 \nu a} \Big[ -64\omega^2(a^2 - b^2)^2(\omega^2
+ \nu^2) A^2_{\psi} \cr & \cr && -32\omega^3 b \kappa^2(a^2 -
b^2)A_{\psi} - 4\omega^2 b^2 \kappa^4 + a^2 \nu^2[(1 +
\frac{b^2}{a^2})\kappa^2 - \nu^2]^2 \Big]^{\frac{1}{2}},
\end{eqnarray}
and $\nu_1 = \nu_2 = \nu$, $A_2 = A_1 = A$.

The deficit angle is defined as $\delta\phi = \int
\frac{\partial\phi}{\partial y}dy$. There is no change in deficit
angles as the calculation depends on $\xi_y$ which is derived from
virasoro constraints. So they remain unchanged and can be read of
from \cite{Giardino:2011dz}.
\subsection{Some Cases}

Here, we studied some cases where momenta and deficit angles are
large or finite. \noindent
{\underbar{\bf{Case I:}}}\\
 Let us take momenta to be large where as deficit angles are finite.
Here we get the relation as
\begin{equation}
E - J_{\psi} = - \left(T+\frac{B}{2\pi}\right)\sin{\delta\psi} ,
\end{equation}
where $\nu = 0$ and $J_{\psi} = \frac{\omega E}{\kappa} -
\left(T+\frac{B}{2\pi}\right)\cos{2\xi_{+}}$. The dispersion
relation is
\begin{equation}
\sqrt{E^2 - {J_{\phi}}^2} - J_{\psi} = - \Big(T +
\frac{B}{2\pi}\Big)\sin{\left(\sqrt{1 +
\frac{4J^2}{T^2{\pi}^2}}~\delta\psi\right)} , \nonumber \\
\label{25}
\end{equation}
where $J_{\phi} = J_{\phi_1} + J_{\phi_2}$,~ $J = J_{\phi_1}
-J_{\phi_2}$,~ $\sin{2\xi_{+}} = \frac{ b }{ a }\sqrt{1 +
\frac{{\nu}^2}{{\omega}^2}}$, ~ $\frac{\nu}{\omega} =
\frac{2J}{T\pi} .$

\noindent {\underbar{\bf{Case II:}}}\\
Here we take momenta associated with $\phi_i$ coordinates as large
and finite value for momentum with $\psi$. We get the dispersion
relation
\begin{equation}
\sqrt{E^2 - {J_{\phi}}^2} + T \delta \psi = \frac{\kappa^2 -
\nu^2}{8\omega^2 J} \pi T^2 (\delta\phi_1 + \delta \phi_2),
\nonumber \\ \label{26}
\end{equation}
where
\begin{equation}
\delta\psi = -\frac{E}{T}\frac{1}{\sqrt{1+\frac{{\nu}^2{ b }^2}{{
a }^2{\omega}^2}}} + \frac{{ a }^2\omega}{4\nu{ b
}^2}(\delta\phi_1 + \delta\phi_2),
\end{equation}
\begin{equation}
J_{\psi} = -\left(T + \frac{B}{2\pi}\right)\cos{2\xi_{+}}.
\end{equation}

\noindent {\underbar{\bf{Case III:}}}\\
Here $J_{\phi_i}$'s are finite where as $J_{\psi}$ is large. The
dispersion relation is found to be
\begin{equation}
\frac{E}{\sqrt{1 + \frac{{ a }^2}{{ b
}^2}\frac{4J^2}{T^2{\pi}^2}}} - J_{\psi} = \left(T +
\frac{B}{2\pi}\right)\cos{2\xi_{+}}, \nonumber \\ \label{27}
\end{equation}
where
\begin{equation}
J_{\psi} = \frac{\omega E}{\kappa} - \left(T +
\frac{B}{2\pi}\right)\cos{2\xi_{+}}.
\end{equation}

We can notice that the two spin string solutions in (\ref{15}) and
(\ref{16}) with $J_{\phi}$ tending to zero give us the one spin
giant magnon and a single spike solution. But in case of three
spin solutions in (\ref{25}), (\ref{26}) and (\ref{27}), it gives
us a new kind of solution.

\section{Folded strings}
In this section, we wish to study a class of folded string
solution in the background of AdS$_4 \times$ CP$^3$ in the
presence of a B-field. We take the ansatz such that $t, \phi,
\psi, \phi_1, \phi_2$ are functions of $\tau$ only and $\rho,
\theta, \xi, \theta_1, \theta_2$ are periodic functions of
$\sigma$ only.
Then the Polyakov action of the string reads as
\begin{eqnarray}
S &=& T \int~d\sigma d\tau \Big[ \frac{1}{4}\cosh^2\rho{\dot{t}}^2
+ \frac{1}{4}\sinh^2\rho({\theta^{\prime}}^2 - \sin^2\theta
\dot{\phi}^2) + {\xi^{\prime}}^2 \cr & \cr &&- \cos^2\xi\sin^2\xi(
\dot{\psi} + \frac{1}{2}\cos{\theta_1}\dot{\phi_1} -
\frac{1}{2}\cos{\theta_2}\dot{\phi_2})^2 + \frac{1}{4}
\cos^2\xi({\theta_1^{\prime}}^2 - \sin^2\theta_1\dot{\phi_1}^2)
\cr & \cr &&+ \frac{1}{4}\sin^2\xi({\theta_2^{\prime}}^2 -
\sin^2\theta_2\dot{\phi_2}^2)\Big] + \frac{B}{4\pi}\int~d\sigma
d\tau \Big[ \sin{2\xi} {\xi}^{\prime}( 2\dot{\psi} +
\cos{\theta_1} \dot{\phi_1} - \cos{\theta_2}\dot{\phi_2}) \cr &
\cr &&+ \cos^2\xi \sin{\theta_1} {\theta_1}^{\prime} \dot{\phi_1}
+ \sin^2\xi \sin{\theta_2} {\theta}^{\prime} \dot{\theta_2} \Big]
,
\end{eqnarray}
The equation of motions for $\rho, \theta, \xi, \theta_1,
\theta_2$ are,
\begin{eqnarray}
{\rho}^{\prime\prime} &=& \sinh\rho\cosh\rho ({\dot{t}}^2 -
\sin^2\theta {\dot{\phi}}^2 + {\theta^{\prime}}^2), \cr & \cr
\theta^{\prime\prime} &=& \sin\theta\cos\theta - 2\coth\rho
\theta^{\prime}\rho^{\prime}, \cr & \cr \xi^{\prime\prime} &=& -
\frac{1}{4}\sin 4\xi(\dot{\psi} + \frac{1}{2}
\cos\theta_1\dot{\phi_1} - \frac{1}{2} \cos\theta_2
\dot{\phi_2})^2 + \frac{1}{4} \cos\xi\sin\xi (\sin^2\theta_1
{\dot{\phi_1}}^2 - \sin^2\theta_2 {\dot{\phi_2}}^2 -
{\theta_2^{\prime}}^2), \cr & \cr \theta_1^{\prime \prime} &=& 2
\sin^2\xi\sin\theta_1(\dot{\psi} + \frac{1}{2}
\cos\theta_1\dot{\phi_1} - \frac{1}{2} \cos\theta_2
\dot{\phi_2})\dot{\phi_1} -
\sin\theta_1\cos\theta_1{\dot{\phi_1}}^2 + 2 \tan\xi \xi^{\prime}
{\theta_1}^{\prime},\cr & \cr \theta_2^{\prime\prime} &=& - 2
\sin^2\xi\sin\theta_2(\dot{\psi} + \frac{1}{2}
\cos\theta_1\dot{\phi_1} - \frac{1}{2} \cos\theta_2
\dot{\phi_2})\dot{\phi_2} -
\tan^2\xi\sin\theta_2\cos\theta_2{\dot{\phi_2}}^2 \cr & \cr &&+ 2
\tan\xi \xi^{\prime} {\theta_2}^{\prime}.
\end{eqnarray}
Choosing $ t = \kappa \tau, \phi = \nu\tau, \psi = \omega\tau,
\phi_{i=1,2} = \nu_i \tau $, we get the conserved quantities as
\begin{equation}
E = \frac{T}{2} \pi \kappa \cosh^2\rho,
\end{equation}
\begin{equation}
J_{\phi} = \frac{T}{2}\int~d\sigma ~\nu \sinh^2\rho ~\sin^2\theta,
\end{equation}
\begin{equation}
J_{\psi} = 2T \int ~ d\sigma \cos^2\xi~\sin^2\xi(\omega +
\frac{1}{2} \nu_1\cos{\theta_1} - \frac{1}{2}\nu_2\cos{\theta_2} )
+ \frac{B}{2\pi} \int~ d\sigma~{\xi}^{\prime} \sin{2\xi},
\end{equation}
\begin{eqnarray}
J_{\phi_1} &=& 2T \int ~ d\sigma~ \Big[ \frac{1}{4} \nu_1
\cos^2\xi \sin^2\theta_1  + \cos^2\xi \sin^2\xi( \frac{1}{2}
\omega\cos\theta_1 + \frac{1}{4} \nu_1 \cos^2\theta_1 \cr& \cr &-&
\frac{1}{4}\nu_2\cos\theta_1 \cos\theta_2)\Big] + \frac{B}{4\pi}
\int ~ d\sigma \Big[
 \sin{2\xi} \cos\theta_1 {\xi}^{\prime} + \cos^2\xi \sin\theta_1
 {\theta_1}^{\prime}\Big],
\end{eqnarray}
\begin{eqnarray}
J_{\phi_2} &=& 2T \int ~ d\sigma \Big[ \cos^2\xi~\sin^2\xi(-
\frac{1}{2} \omega \cos\theta_2 - \frac{1}{4} \nu_1 \cos\theta_1
\cos\theta_2 + \frac{1}{4} \nu_2 \cos^2\theta_2) \cr & \cr &+&
\frac{1}{4} \nu_2 \sin^2\xi \sin^2\theta_2 \Big] + \frac{B}{4\pi}
\int~ d\sigma \big[ \sin^2\xi \sin\theta_2 {\theta_2}^{\prime} -
\sin 2\xi \cos\theta_2 {\xi}^{\prime} \Big].
\end{eqnarray}
The Virasoro constraints remain unchanged as in
\cite{Chen:2008qq}. For folded string we choose ansatz as
$\theta_1 = \theta_2 =0$. Then the angular momenta changed to
\begin{eqnarray}
J_{\psi} &=& 2T \int~ d\sigma~\Theta\cos^2\xi \sin^2\xi +
\frac{B}{2\pi} \int~d\sigma~\sin 2\xi~ \xi^{\prime} ,\cr & \cr
J_{\phi_1} &=& T \int~ d\sigma~\Theta\cos^2\xi \sin^2\xi +
\frac{B}{4\pi} \int~d\sigma~\sin 2\xi~ \xi^{\prime} ,\cr & \cr
J_{\phi_2} &=& -T \int~ d\sigma~\Theta\cos^2\xi \sin^2\xi -
\frac{B}{4\pi} \int~d\sigma~\sin 2\xi~ \xi^{\prime},
\end{eqnarray}
where $\Theta = \omega + \frac{1}{2}(\nu_1 - \nu_2)$ and this
gives
\begin{equation}
J_{\psi} = 2T \big[K(q) - E(q)] + \frac{B}{2\pi} \sin^2 \xi_0,
\end{equation}
where K(q) and E(q) are complete elliptic integral of the first
and second kind with $q =  \sin 2\xi_0$. At $\xi_0 =
\frac{\pi}{4}$ , both E and J's diverge but gives the dispersion
relation
\begin{equation}
E - J_{\psi} = 2T - \frac{B}{4\pi}.
\end{equation}
\section{Conclusions} In this paper we have studied several rigidly
rotating string solutions in the background of AdS$_4 \times$
CP$^3$ in the presence of B$_{\rm NS}$ holonomy. First we have
studied a class of solutions corresponding to the so called giant
magnon solution with two spins along a subspace of CP$^3$. The
corresponding dispersion relation among various charges gets
contribution from the B-field. Then we have generalized to three
spin solutions in another subspace of CP$^3$. We have studied
various limiting cases of the solutions and have written down the
corresponding dispersion like relations. The finite size
corrections to the dispersion relations can be calculated by
following \cite{Arutyunov:2006gs}. The dual field theory results
for the present class of solutions is not completely understood,
even though the ABJ theory was shown to be integrable and all
string spectrum was unaffected by the presence of discrete B$_{\rm
NS}$ holonomy \cite{Bak:2008vd}. Further it was shown that the
spectrum of all single trace operators is independent of B$_{\rm
NS}$ holonomy as well. Hence it appears that the dual operators
corresponding to the class of solutions presented here are
tractable. The solutions presented here are different from the
ones presented earlier is due to the boundary conditions used
here. We have used open string boundary condition which correspond
to open spin chain. This fact is essentially responsible for
giving a $B$-dependent term in the dispersion relation among
various charges as suggested in \cite{Bak:2008vd} from the
standpoint of the dual gauge theory. Hence it would really be
interesting to investigate the gauge theory dual of the solutions
presented here as to see whether the boundary operators gets
corrected due to this.

\end{document}